\documentclass{article}

\usepackage[final]{neurips_2024}
\usepackage{amsmath}
\usepackage{graphicx}
\usepackage{algorithm}

\usepackage[utf8]{inputenc} 
\usepackage[T1]{fontenc}    
\usepackage{hyperref}       
\usepackage{url}            
\usepackage{booktabs}       
\usepackage{amsfonts}       
\usepackage{nicefrac}       
\usepackage{microtype}      
\usepackage{xcolor}         

\title{From Rattle to Roar: Optimizer Showdown for MambaStock on S\&P 500}

\begin{document}

\author{%
  Maria Garmonina \\
  Department of Applied Mathematics \\
  Columbia University \\
  New York, USA \\
  \texttt{mkg2169@columbia.edu} \\
  \And
  Alena Chan \\
  Department of Computer Science \\
  Columbia University \\
  New York, USA \\
  \texttt{ac5477@columbia.edu} \\
}

\maketitle

\begin{abstract}
  We evaluate the performance of several optimizers on the task of forecasting S\&P 500 Index returns with the MambaStock model. Among the most widely used algorithms, gradient-smoothing and adaptive-rate optimizers (for example, Adam and RMSProp) yield the lowest test errors. In contrast, the Lion optimizer offers notably faster training. To combine these advantages, we introduce a novel family of optimizers, Roaree, that dampens the oscillatory loss behavior often seen with Lion while preserving its training speed. Source code can be found at https://github.com/maria-garmonina/snakes-on-trading-floor.
\end{abstract}

\section{Introduction}
Interest in state-space models has surged as a competitive alternative to both simple statistical approaches and resource-intensive transformers. The recently introduced Mamba architecture has demonstrated strong performance on long time-series sequences \cite{gu2024mamba}, making it particularly well-suited for financial data. In this work, we investigate the use of Mamba-inspired selective state-space models for forecasting financial time series. Specifically, we focus on predicting future returns of the S\&P 500 Index with the MambaStock model \cite{shi2024mambastock}. Our primary contribution is a thorough evaluation of several widely adopted optimization algorithms and the introduction of a novel family of optimizers tailored to this task -- with the goal of improving both predictive accuracy and training efficiency on sequential financial data.

\subsection{Problem Statement}

We explore how the choice of optimizer affects both training speed and forecasting accuracy in the MambaStock model. Training any machine learning model is an optimization problem, and a variety of algorithms exist for approximating the optimal parameters. We examine the trade-offs among predictive error, convergence rate, and overall training time -- each metric is critical for practical trading applications.  

\subsection{Objectives and Scope}
This paper has two objectives:
\begin{enumerate}
  \item Compare the performance of the MambaStock model under a range of optimizers.
  \item Leverage the speed vs. MSE trade-off to design and validate a new optimizer family.
\end{enumerate}

\section{Literature Review}

\subsection{Related Work}
Traditional time‐series models such as ARIMA assume linear relationships between past and future values and often fail to capture sudden market shocks, leading to inaccurate price predictions. Hybrid LSTM-CNN architectures have improved sensitivity to market shifts \cite{zhong2021data}, but their computational demands can be prohibitive. Similarly, Kalman filters offer interpretable price trend filtering yet remain constrained by linearity assumptions.

More recently, state-space models (SSMs) have gained traction for their ability to process very long sequences efficiently. Unlike transformers, which require \(O(n^2)\) attention operations and thus struggle with long-history inference (recalculating attention over extended price histories is impractical), SSMs scale linearly with input length and maintain constant memory usage. This makes them well-suited for real-time financial forecasting, especially in resource-constrained environments where training or deploying a full transformer might be prohibitive.

Gu et al. \cite{gu2022s4} introduced the Structured State-Space Sequence (S4) model, which reparameterizes transition matrices to capture long-range dependencies with efficient computation. Building on S4, Gu \& Dao \cite{gu2024mamba} proposed Mamba, adding an optimized recurrent update block scan and a hardware optimization by parallelizing computation via convolution. Mamba’s ability to vary its SSM parameters based on the input enables it to focus on relevant information and forget the rest. Thus, Mamba often matches the accuracy of attention-based models while offering linear time complexity and fast inference.

\subsection{Identification of Gaps in Existing Research}
For MambaStock \cite{shi2024mambastock}, Shi adapted the Mamba architecture to the stock price prediction task. It uses historical series data to predict future price. While this model brought gains in accuracy, further optimizations of this model have not been studied much. 

While the Lion optimizer (Chen et al., 2023) \cite{chen2023lion} has emerged as a memory-efficient alternative to Adam-style methods by leveraging a sign-based momentum, optimizers specifically designed for financial time series within Mamba architectures remain largely unexplored -- most prior work has focused on transformer-based models.

\section{Methodology}

\subsection{Data Collection and Preprocessing}
The dataset contains weekly S\&P 500 Index observations from 2000 to 2019 with historical returns, 10 engineered technical‑analysis signals, three fundamental valuation ratios, and two crowd‑sentiment scores (Table \ref{tab:features}) -- collected by Zhong et al. \cite{zhong2021data}. The prediction target is the forward one‑week return (\texttt{return\_t\_plus\_1}), obtained by shifting the adjusted closing price one week ahead.

\begin{table}[ht]
\centering
\label{tab:features}
\begin{tabular}{lp{10cm}}
\toprule
\textbf{Attribute} & \textbf{Description} \\
\midrule
\multicolumn{2}{l}{\textit{Target}} \\
return\_t\_plus\_1 & Forward one-week return: \((\texttt{adj\_close}_{t+1} / \texttt{adj\_close}_t) - 1\) \\
\midrule
\multicolumn{2}{l}{\textit{History}} \\
return\_t & Return for the current week \\
\midrule
\multicolumn{2}{l}{\textit{Technical Indicators}} \\
adx            & Average Directional Index -- measures trend strength (higher = stronger trend) \\
adxr           & Average Directional Movement Rating -- a smoothed version of ADX \\
trix           & Triple Exponential Average -- a moving average of a moving average of a moving average, capturing momentum \\
cci            & Commodity Channel Index -- identifies overbought/oversold conditions \\
macdh          & MACD Histogram -- the difference between the MACD line and its signal line \\
rsi\_14        & Relative Strength Index (14‐day) -- measures recent gains vs. losses (RSI > 70 overbought, < 30 oversold) \\
kdjk           & K‐line of the Stochastic Oscillator -- another momentum indicator \\
wr\_14         & Williams \%R (14‐day) -- overbought/oversold oscillator \\
atr            & Average True Range -- measures volatility \\
atr\_percent   & ATR as a percentage of price -- normalized volatility measure \\
\midrule
\multicolumn{2}{l}{\textit{Valuation Ratios}} \\
PbRatio        & Price‐to‐Book Ratio -- price divided by book value per share \\
PeRatio        & Price‐to‐Earnings Ratio -- price divided by earnings per share \\
PsRatio        & Price‐to‐Sales Ratio -- price divided by revenue per share \\
\midrule
\multicolumn{2}{l}{\textit{Sentiment Scores}} \\
spsentiment    & Crowd sentiment (e.g., from Seeking Alpha, StockTwits) \\
sentiment      & General daily sentiment score (positive/negative/neutral) \\
\bottomrule
\end{tabular}
\caption{Summary of features in the weekly S\&P 500 Index dataset}
\end{table}

We split the dataset into train, validation, and test parts: the test set contains the last 100 weeks of observations, the training set is 90\% of the remaining data, with 10\% being the validation data (used for tracking the validation loss over the training epochs). We keep all of our datasets causal -- not letting the model peek into the future.

\subsection{Model Selection}
We utilized the MambaStock model, which has proven itself for successful Chinese stock market predictions, with hidden size set to 64 and 2 layers. No architectural
changes were made -- the optimizer choice is the sole experimental factor.

\subsection{Baseline Optimizers}

We benchmark eight widely‑used methods: SGD, SGD with Momentum, Nesterov, RMSProp, Adagrad, Adam, AdamW, and Lion.

\subsection{Roaree: a Smooth‑Lion Family}
Due to the computational efficiency of the Lion optimizer, we based our Roaree algorithms on Lion. Lion’s original update
$
\theta_{t}=\theta_{t-1}-\eta_t\ (\operatorname{sign}(c_t) + \lambda \theta_{t-1})\ , \
$$
c_t=\beta_2 m_{t-1}+(1-\beta_2)g_t
$
is \emph{non‑differentiable}.  
Roaree replaces \(\operatorname{sign}(\cdot)\) with a
\emph{smooth surrogate} $s_{\kappa}(\cdot)$ controlled by a
\textit{curvature} hyper‑parameter $\kappa>0$ (please refer to Algorithm 1):

\begin{align}
s\_{\kappa}^{\text{tanh}}(x)      &= \tanh(\kappa x)\\
s\_{\kappa}^{\text{atan}}(x)      &= \tfrac{2}{\pi}\arctan(\kappa x)\\
s\_{\kappa}^{\text{softsign}}(x)  &= \frac{\kappa x}{1+|\kappa x|}\\
s\_{\kappa}^{\text{sigmoid}}(x)   &= 2\sigma(\kappa x)-1\\
s\_{\kappa}^{\text{erf}}(x)       &= \operatorname{erf}(\kappa x)\\
s\_{\kappa}^{\text{norm}}(x)      &= \frac{x}{\sqrt{x^{2}+1}}\quad
\end{align}

\begin{figure}[hbt!]
    \centering
    \includegraphics[scale=0.22]{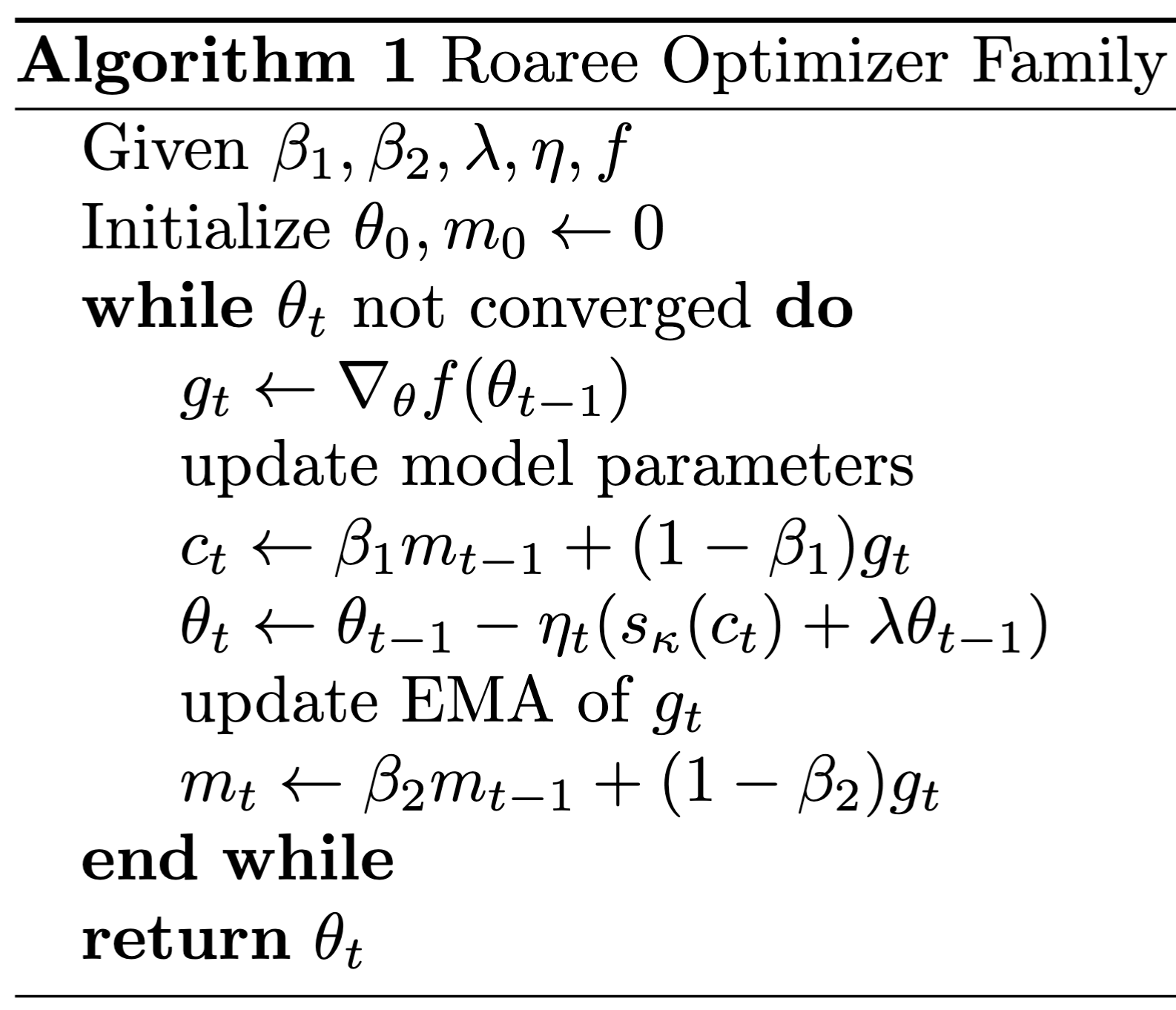}
\end{figure}

Setting $\kappa\!\to\!\infty$ recovers the hard sign, whereas a small $\kappa$ yields a
nearly linear step.

We sweep
\(\kappa\in\{10,\,100,\,1000\}\).

\subsection{Benchmarking Protocol and Evaluation Metrics}
All optimizers were trained for 64 epochs on MambaStock with identical random seeds for reproducibility. 

Standard optimizers were explored on the larger grid:
\[
\;
  \text{lr}\in
  \bigl\{10^{-5},\;5\!\times\!10^{-5},\;10^{-4},\;5\!\times\!10^{-4}, 10^{-3},\;5\!\times\!10^{-3},\; 10^{-2},\;5\!\times\!10^{-2} \bigr\}, \]
  \;
  \[
  \text{wd}\in
  \bigl\{0, 10^{-4},\;5\!\times\!10^{-4},\; 10^{-3},\;5\!\times\!10^{-3},\;10^{-2},\;5\!\times\!10^{-2},\;10^{-1}\bigr\}.
\]

For the Roaree family we narrowed the grid on
learning rate and weight decay in order to explore six surrogates and three curvature values:

\[
\;
  \text{lr}\in
  \bigl\{10^{-4},\;10^{-3},\;10^{-2}\bigr\},
  \qquad
  \;
  \text{wd}\in
  \bigl\{0,\;10^{-3},\;10^{-2},\;10^{-1}\bigr\}.
\]

For every configuration we log the following metrics:

\begin{itemize}
    \item Average Epoch Time -- wall‑clock seconds per epoch
    \item MSE / RMSE / MAE -- scale‑dependent error measures
    \item R\textsuperscript{2} -- explained variance
    \item Directional Accuracy -- gauges whether the model correctly predicts the direction of change
\end{itemize}

Training and validation loss history (MSE) over the epochs is collected for convergence analysis. The test set is reported once per optimizer at the final configuration.

\section{Experimental Results}

\subsection{Experimental Setup}
\begin{itemize}
    \item Hardware: NVIDIA T4 GPU in Google Colab \
    \item Software: MambaStock, \texttt{torch.optim}, \texttt{pytorch-optimizer}
\end{itemize}

\subsection{Results and Analysis}

\subsubsection{Baseline Optimizers}

\begin{figure}[hbt!]
    \centering
    \includegraphics[scale=0.5]{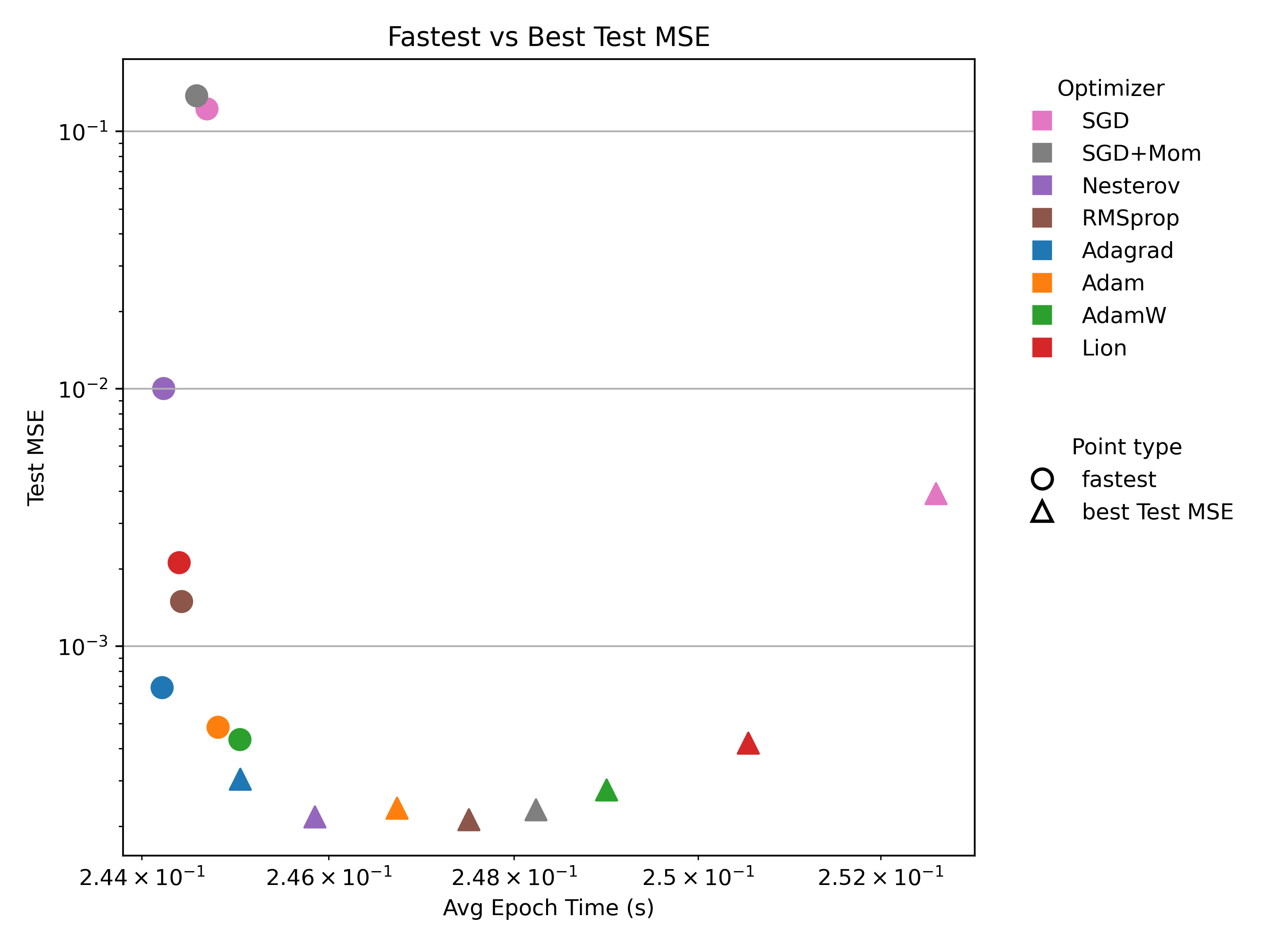}
    \caption{Best Speed vs. Best Test MSE}
    \label{fig:mse-baseline}
\end{figure}

Across our baseline optimizers, we observe the lowest test errors with Nesterov, RMSProp, Adam, and SGD with momentum (Fig. \ref{fig:mse-baseline}). Because financial-return targets are very small and noisy, gradients can vary by orders of magnitude across the layers of a Mamba block. Optimizers that smooth gradients over time (momentum) and/or adapt the learning rate per parameter (RMS-style methods) handle this variability better than either Lion or vanilla SGD.

Notably, AdamW -- extensively adopted for transformer training -- does not rank among the top performers for our task. Decoupled weight decay is most beneficial when you need strong regularization, but here the signal is already vanishingly small, so extra shrinkage slows convergence. Adam (without separate decay) still applies adaptive moments yet avoids the over-regularization, landing in a lower-error basin.

\begin{figure}[hbt!]
    \centering
    \includegraphics[scale=0.3]{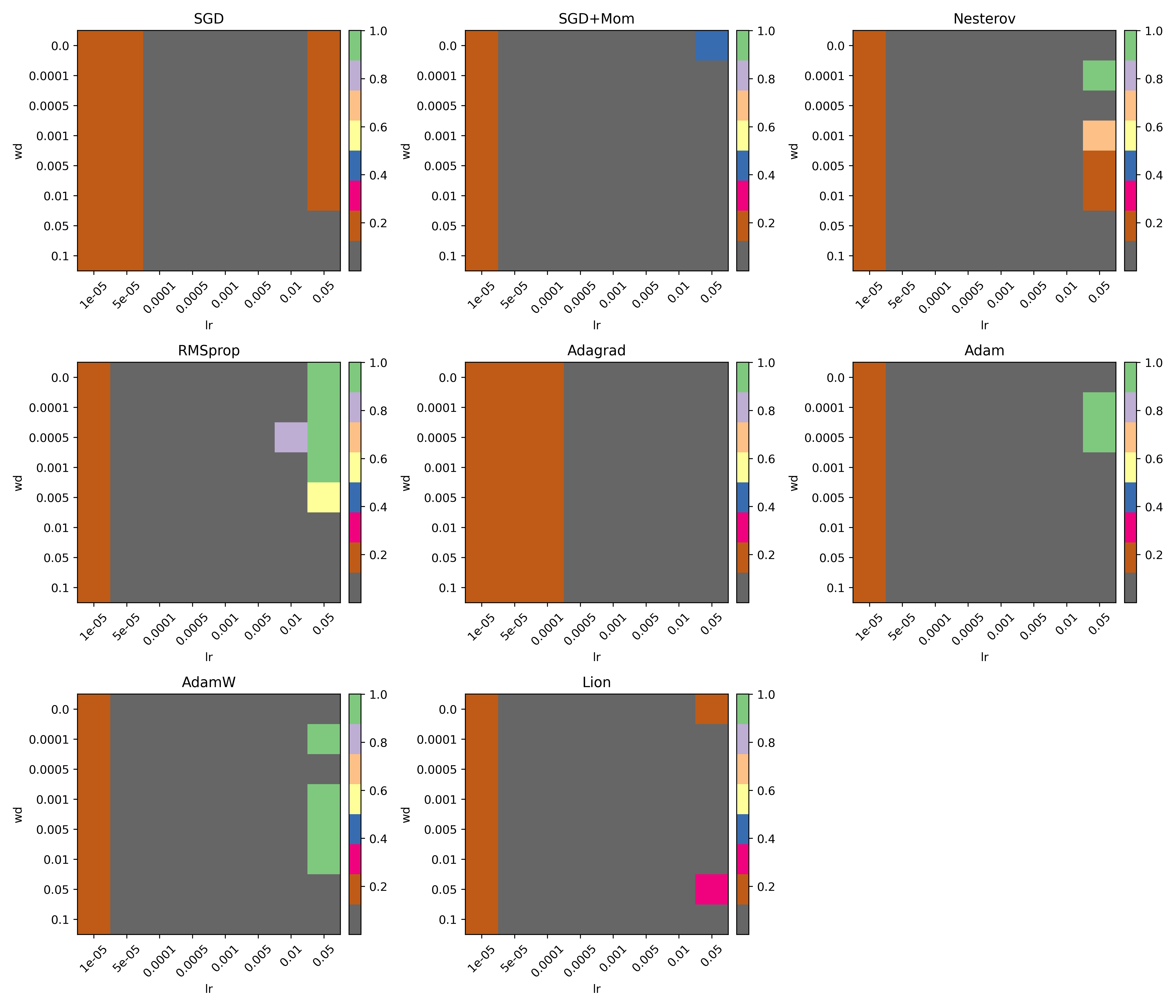}
    \caption{Validation MSE Heatmaps}
    \label{fig:heatmaps}
\end{figure}

We also find that Lion supports a broader range of learning rates and weight-decay settings for which validation MSE remains low (Fig. \ref{fig:heatmaps}). So in scenarios requiring aggressive hyperparameters -- such as large-scale training or extensive task exploration -- Lion is the more robust choice. However, this comes with the usual trade-off: although Lion has one of the fastest epoch times, it does not always achieve the absolute lowest test error.

\subsubsection{Roaree Optimizers}
To ensure a fair comparison, we limit all baseline optimizers to the smaller hyperparameter grid used for our Roaree experiments.

\begin{figure}[hbt!]
    \centering
    \includegraphics[scale=0.5]{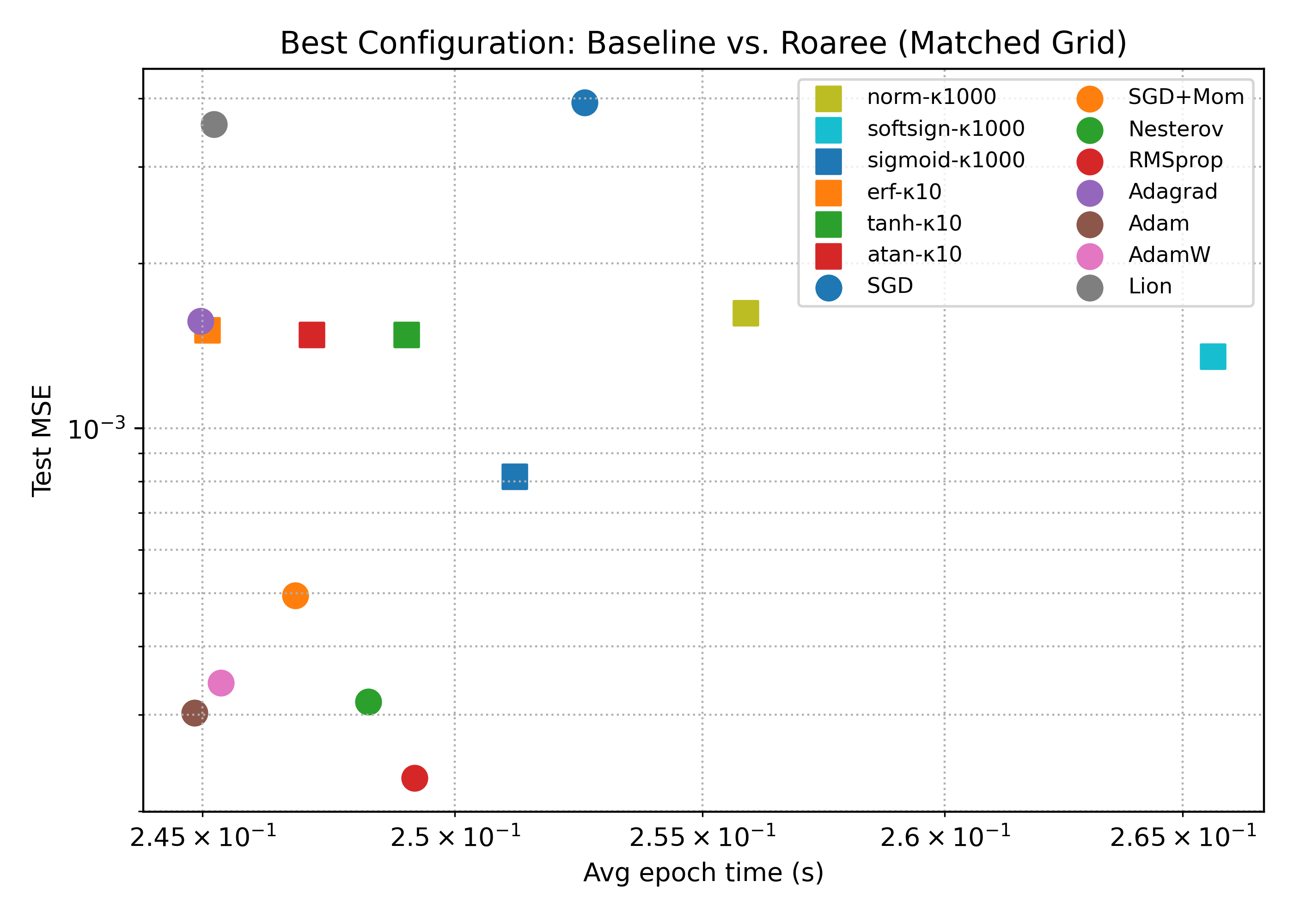}
    \caption{Test MSE: Baseline vs. Roaree}
    \label{fig:mse-roaree}
\end{figure}

\begin{figure}[hbt!]
    \centering
    \includegraphics[scale=0.5]{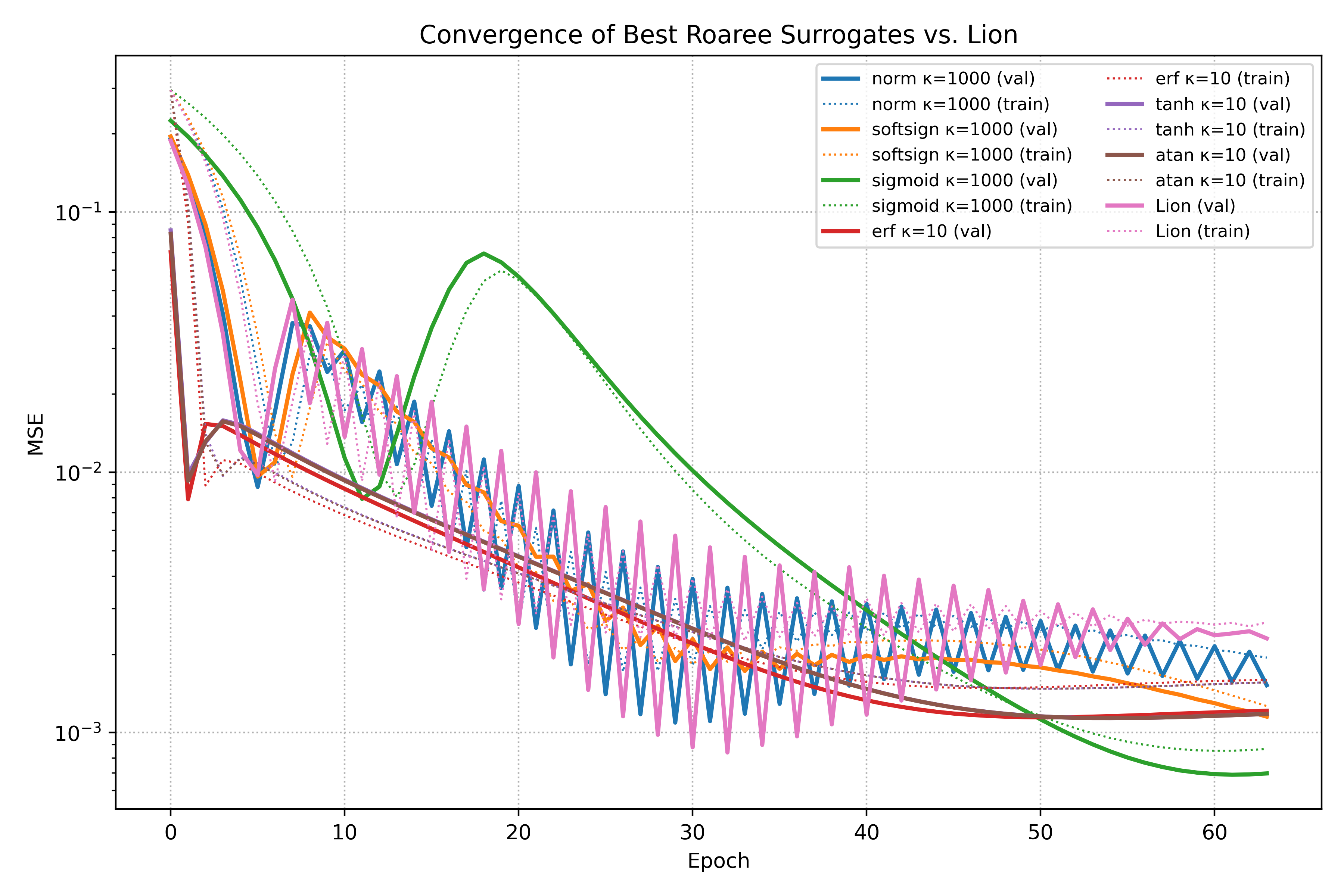}
    \caption{Convergence: Lion vs. Roaree}
    \label{fig:lion-roaree-conv}
\end{figure}

Under these settings, every Roaree variant achieves better accuracy than Lion, but still trails Adam and RMSProp in lowest test MSE (Fig. \ref{fig:mse-roaree}). Moreover, except for the norm surrogate, all smooth‐surrogate algorithms yield markedly smoother convergence -- unlike Lion’s large oscillations (Fig. \ref{fig:lion-roaree-conv}). This shows that replacing the hard sign step with a smooth approximation generally stabilizes training.

In particular, smooth surrogates reduce the bias near the optimum and curb Lion’s overshooting behavior (reflected in the dampened oscillations in Fig. \ref{fig:lion-roaree-conv}). The norm surrogate still exhibits oscillations because its linear region is extremely narrow, but even here the oscillation amplitude is slightly lower than Lion’s.

The best surrogate function choice appears to be $s\_{\kappa}^{\text{erf}}$ with $\kappa=10$. It retains extremely low average epoch time (reducing it even further than Lion) while also decreasing Test MSE.

\section{Further Discussion}

\subsection{Challenges and Limitations}
One major challenge in our analysis was the limited availability of detailed historical stock data, given its commercial value.

A second difficulty lies in comparing optimizers of different natures, since each requires its own hyperparameter regime to achieve peak performance. For example, AdamW typically demands much lower learning rates than Nesterov, and Adagrad benefits from near-zero weight decay while RMSProp often needs higher values. To ensure a fair comparison, we performed a wide‐ranging grid search for each MambaStock and optimizer pairing -- trading off granularity of our experiments.

\subsection{Future Directions}
Second‐order optimizers such as Sophia \cite{liu2023sophia} have demonstrated up to a 2x speed-up over Adam by computing accurate curvature (Hessian) approximations for faster, curvature-aware convergence. In contexts requiring large-scale dataset processing, Sophia could accelerate experimentation with Mamba model variants -- although its complexity demands more extensive hyperparameter tuning than first-order methods.

Similarly, squeezing additional performance from our Roaree family will require deeper grid searches, especially over the curvature parameter $\kappa$ that scales the smooth surrogate horizontally. An optimal $\kappa$ must be large enough to drive rapid learning yet small enough to prevent gradient explosion.

In this work, we evaluated six surrogate functions in place of the hard sign in Lion’s update rule. Future studies might explore alternative smooth sign approximations and refined parameter schedules to further boost convergence and training speed and test accuracy.

\section{Conclusion}

\subsection{Summary of Findings}
In our experiments, the Roaree family of optimizers smooths the convergence behavior of Lion, improving accuracy of predictions. This suggests that smooth sign surrogates allow parameters to reach a more optimal state by stabilizing the training process. The best surrogate sign approximation appears to be $s\_{\kappa}^{\text{erf}}$.

Out of the baseline optimizers, those using adaptive learning rates and momentum achieved the lowest errors.

\subsection{Contributions}
Our main contribution is boosting the performance of the MambaStock model on financial return prediction. We analyze trade-offs between training speed and accuracy achieved by various optimizers, which provides directions for time-sensitive or performance-sensitive training. Furthermore, this comparison guides developers in choosing optimizers for hypothesis and strategy testing. Additionally, we propose a new family of Roaree optimizers that achieve lower errors and more stable convergence than Lion.

Contributions of each team member: 
\begin{itemize}
    \item M. G.: project proposal, initial literature review, data acquisition and preprocessing, Roaree optimizer design, benchmarking loop implementation, data collection, visualizations, paper writeup.
    \item A. C.: literature review, data preprocessing, MambaStock setup, Roaree optimizer design, data collection, paper writeup.
\end{itemize}

\section*{Acknowledgments}
We are grateful to Prof. Richard Zemel for his mentorship in this project.

\section*{Code Availability}
All the code used in this project and the obtained experimental data is available at \url{https://github.com/maria-garmonina/snakes-on-trading-floor.git}.

\raggedbottom

\section*{Appendix}

\begin{figure}[h!]
    \centering
    \includegraphics[scale=0.3]{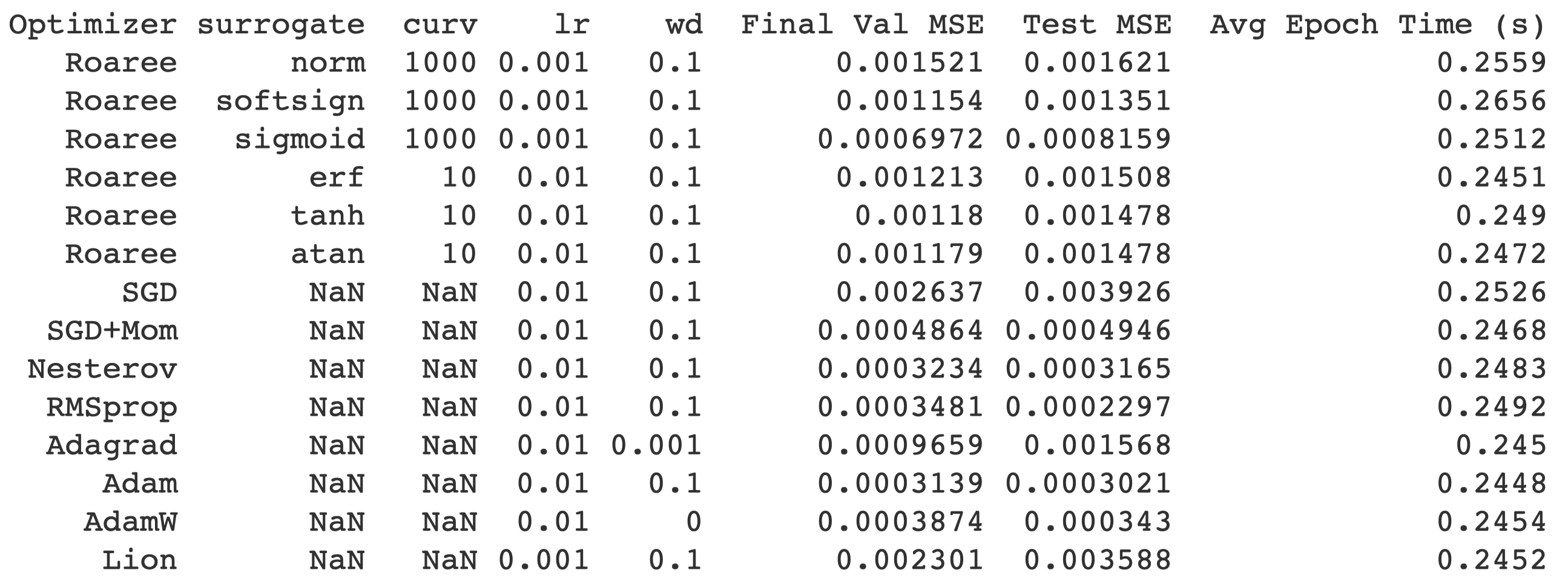}
    \caption{Roaree vs. Baseline: Lowest Errors on the Small Grid}
    \label{fig:numbers}
\end{figure}

\begin{figure}[h!]
    \centering
    \includegraphics[scale=0.5]{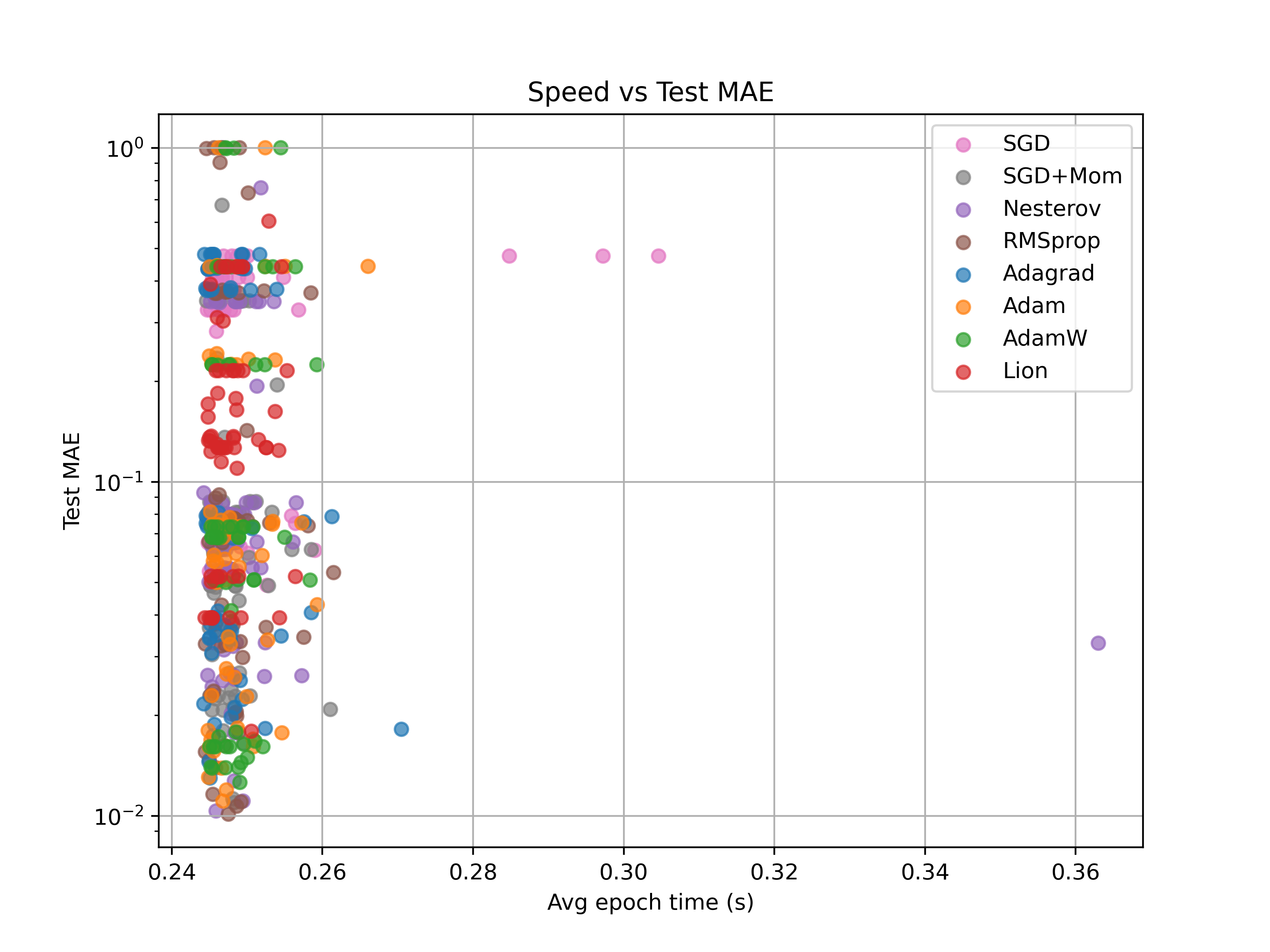}
    \caption{Baseline Optimizers: Speed vs. Test MAE}
    \label{fig:base-opt-MAE}
\end{figure}

\begin{figure}[h!]
    \centering
    \includegraphics[scale=0.5]{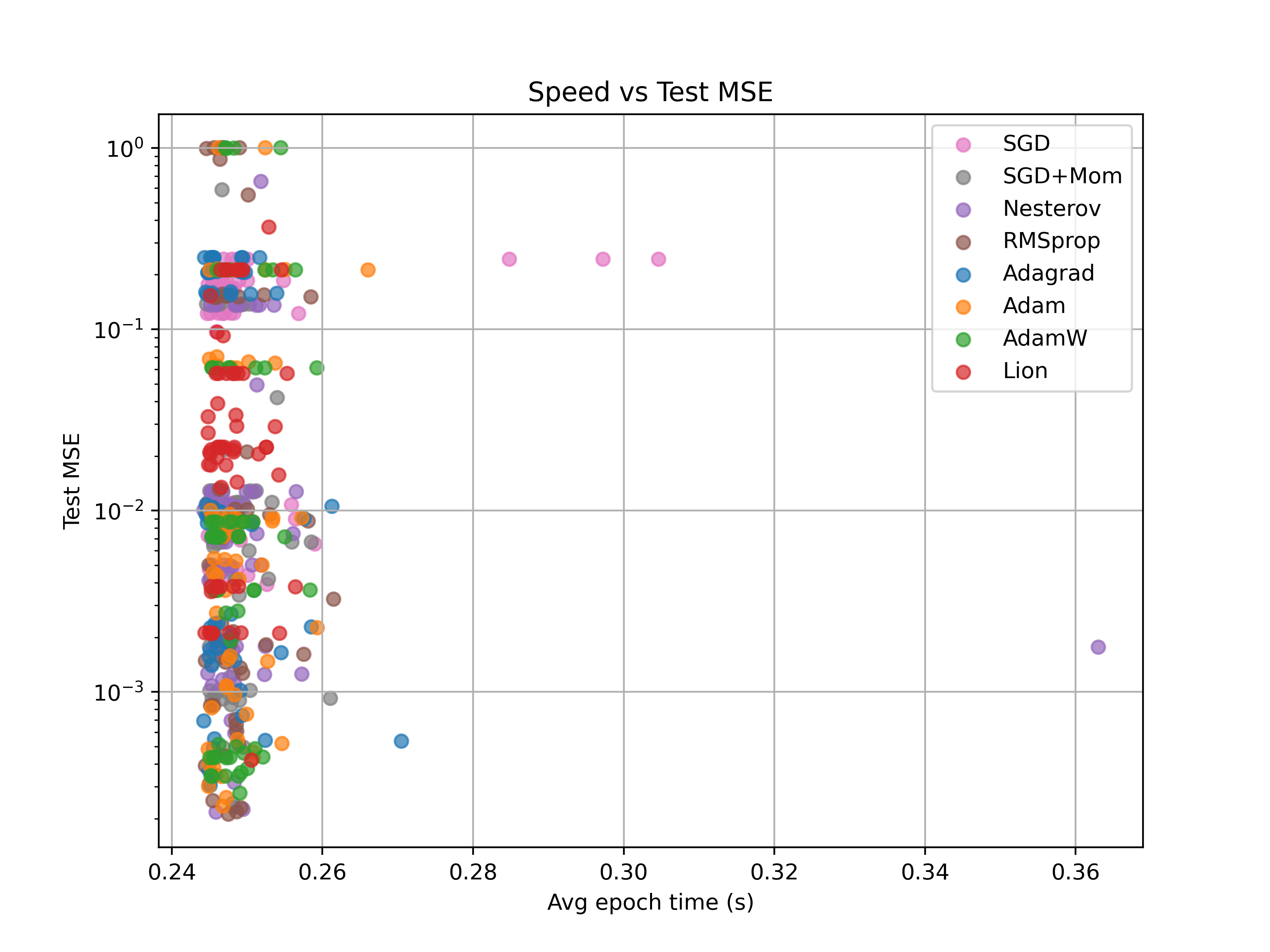}
    \caption{Baseline Optimizers: Speed vs. Test MSE}
    \label{fig:base-opt-MSE}
\end{figure}

\begin{figure}[h!]
    \centering
    \includegraphics[scale=0.5]{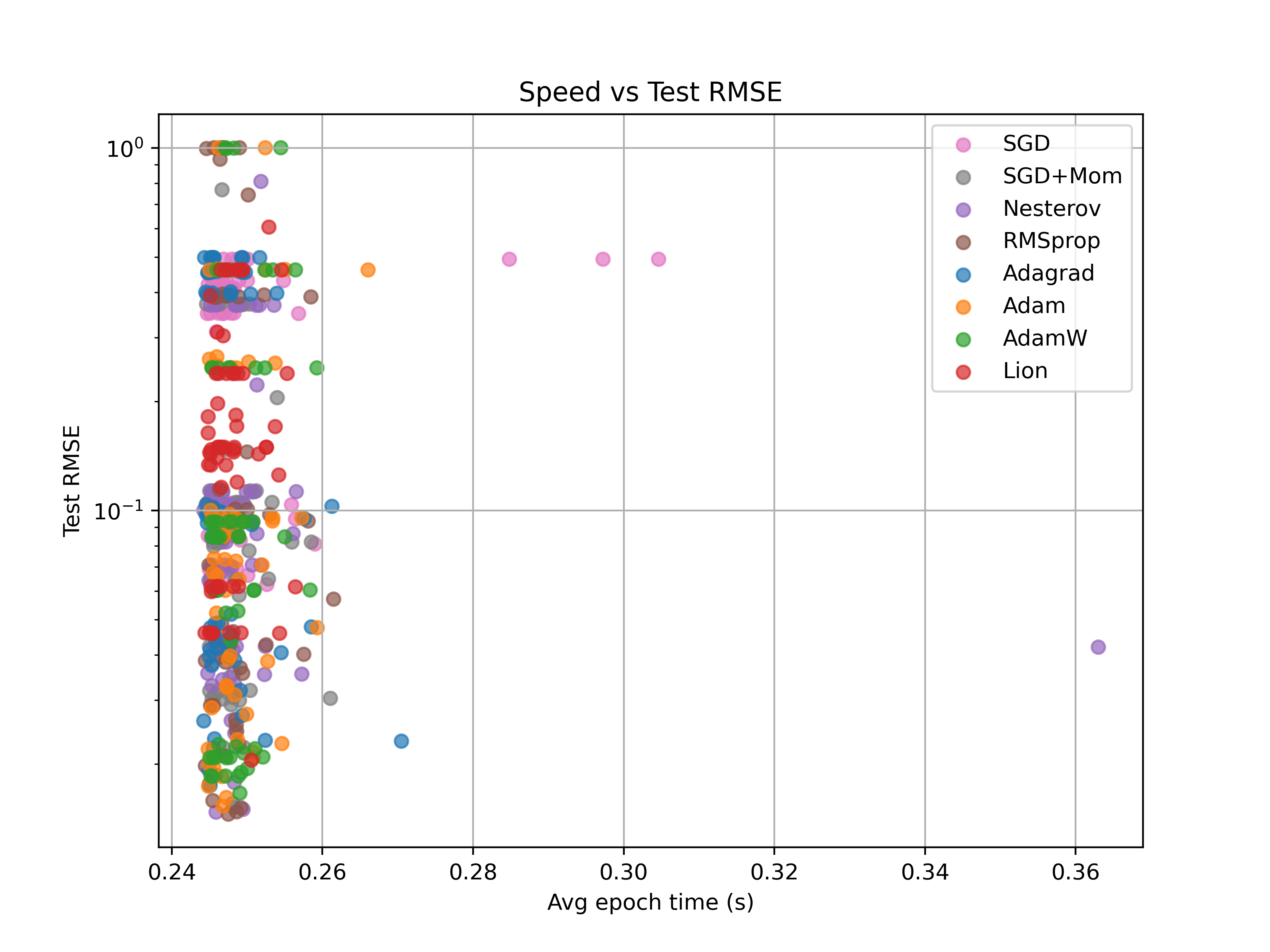}
    \caption{Baseline Optimizers: Speed vs. Test RMSE}
    \label{fig:base-opt-RMSE}
\end{figure}

\begin{figure}[h!]
    \centering
    \includegraphics[scale=0.5]{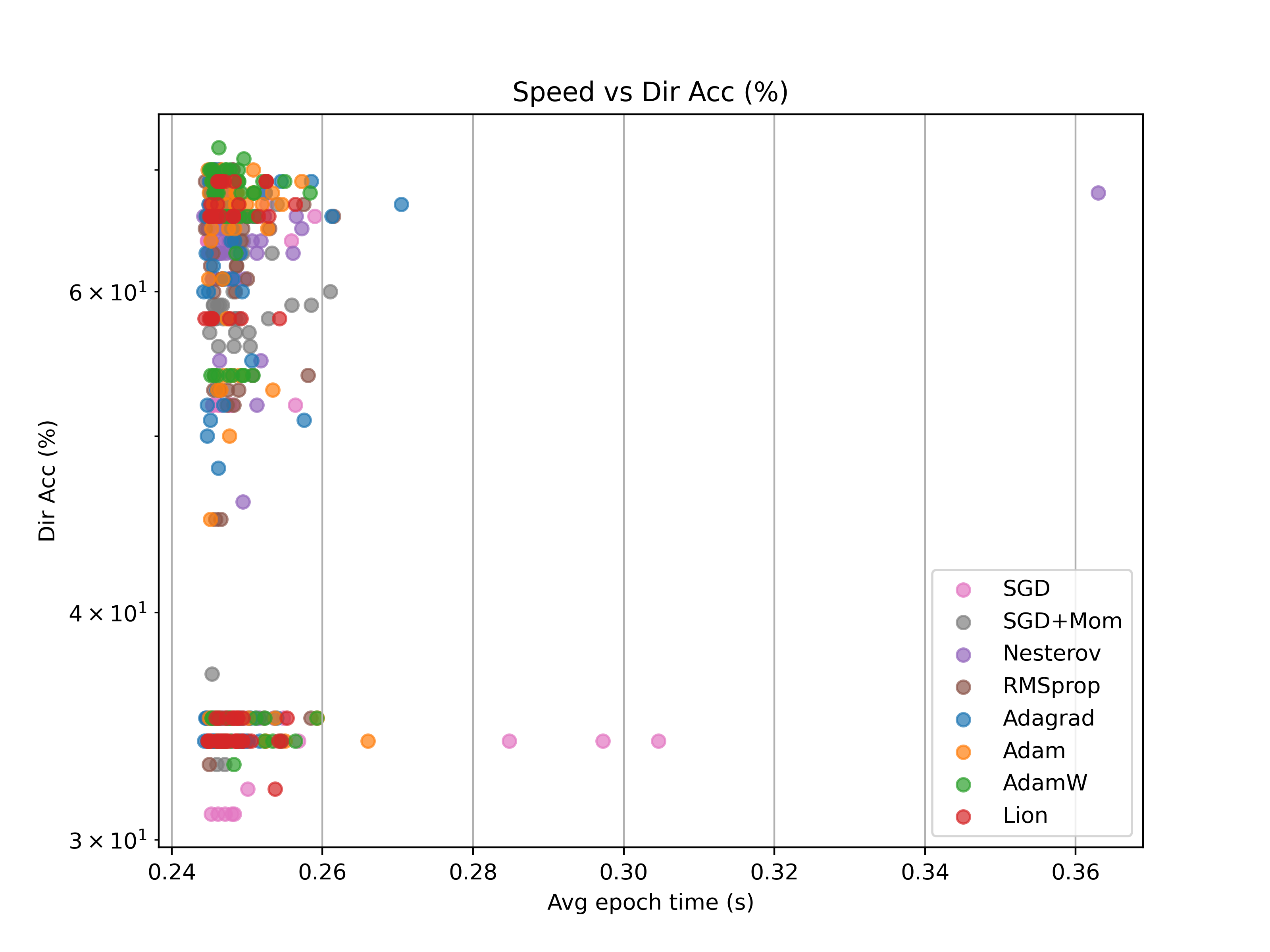}
    \caption{Baseline Optimizers: Speed vs. Directional Accuracy}
    \label{fig:base-opt-dir-acc}
\end{figure}

\begin{figure}[h!]
    \centering
    \includegraphics[scale=0.5]{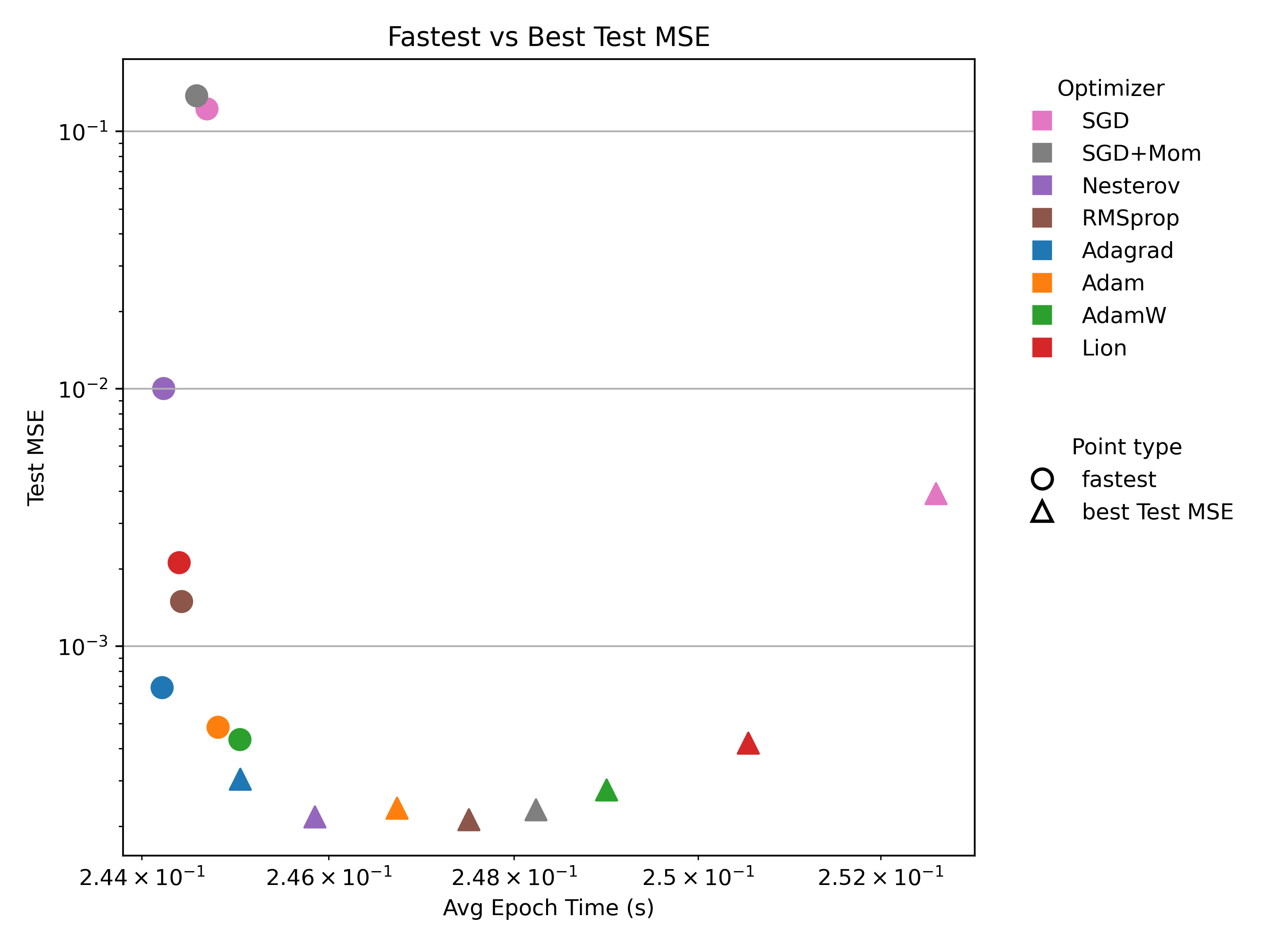}
    \caption{Baseline Optimizers: Fastest vs. Best Test MSE}
    \label{fig:base-opt-best-MSE}
\end{figure}

\begin{figure}[h!]
    \centering
    \includegraphics[scale=0.5]{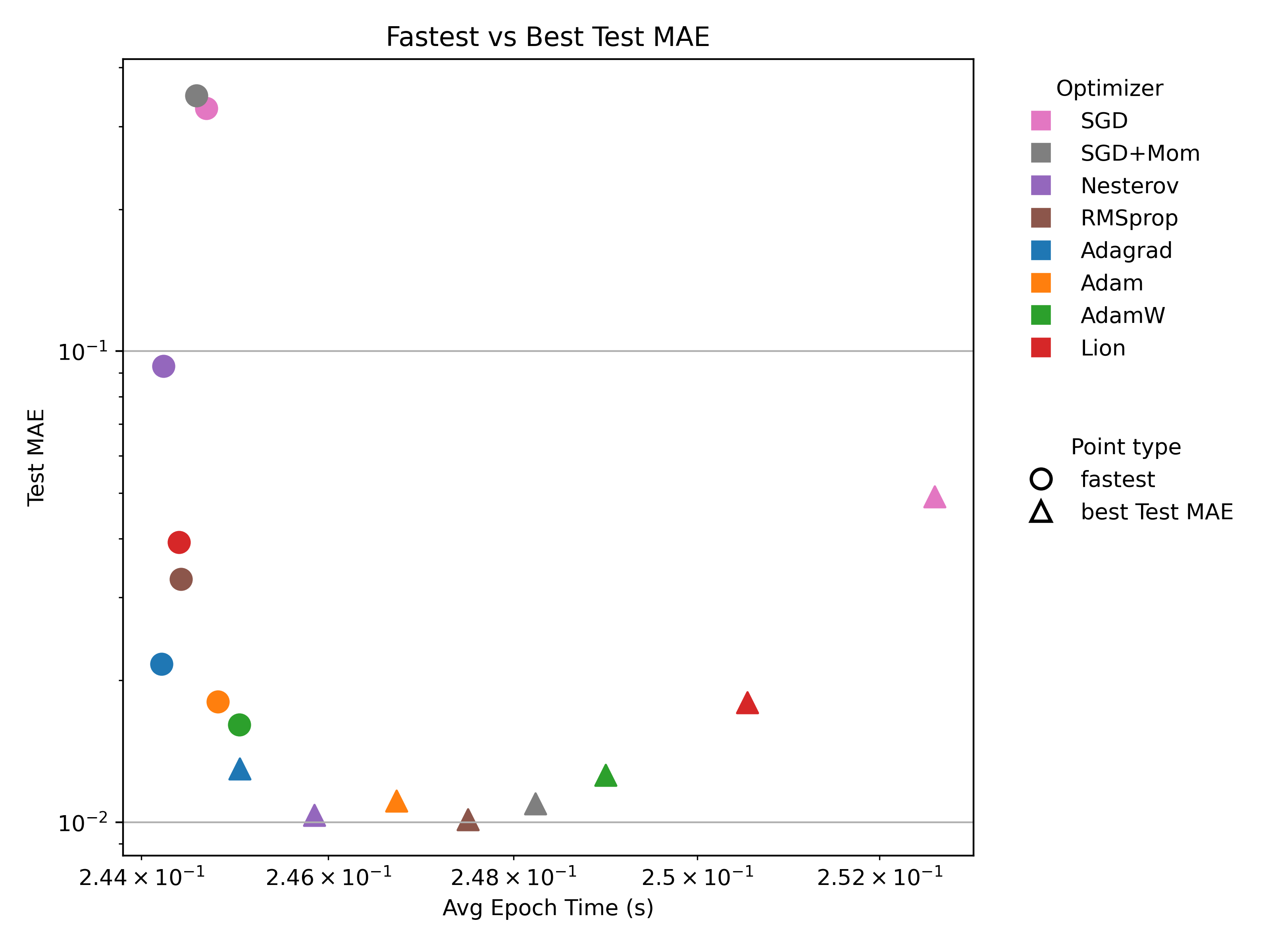}
    \caption{Baseline Optimizers: Fastest vs. Best Test MAE}
    \label{fig:base-opt-best-MAE}
\end{figure}

\begin{figure}[h!]
    \centering
    \includegraphics[scale=0.5]{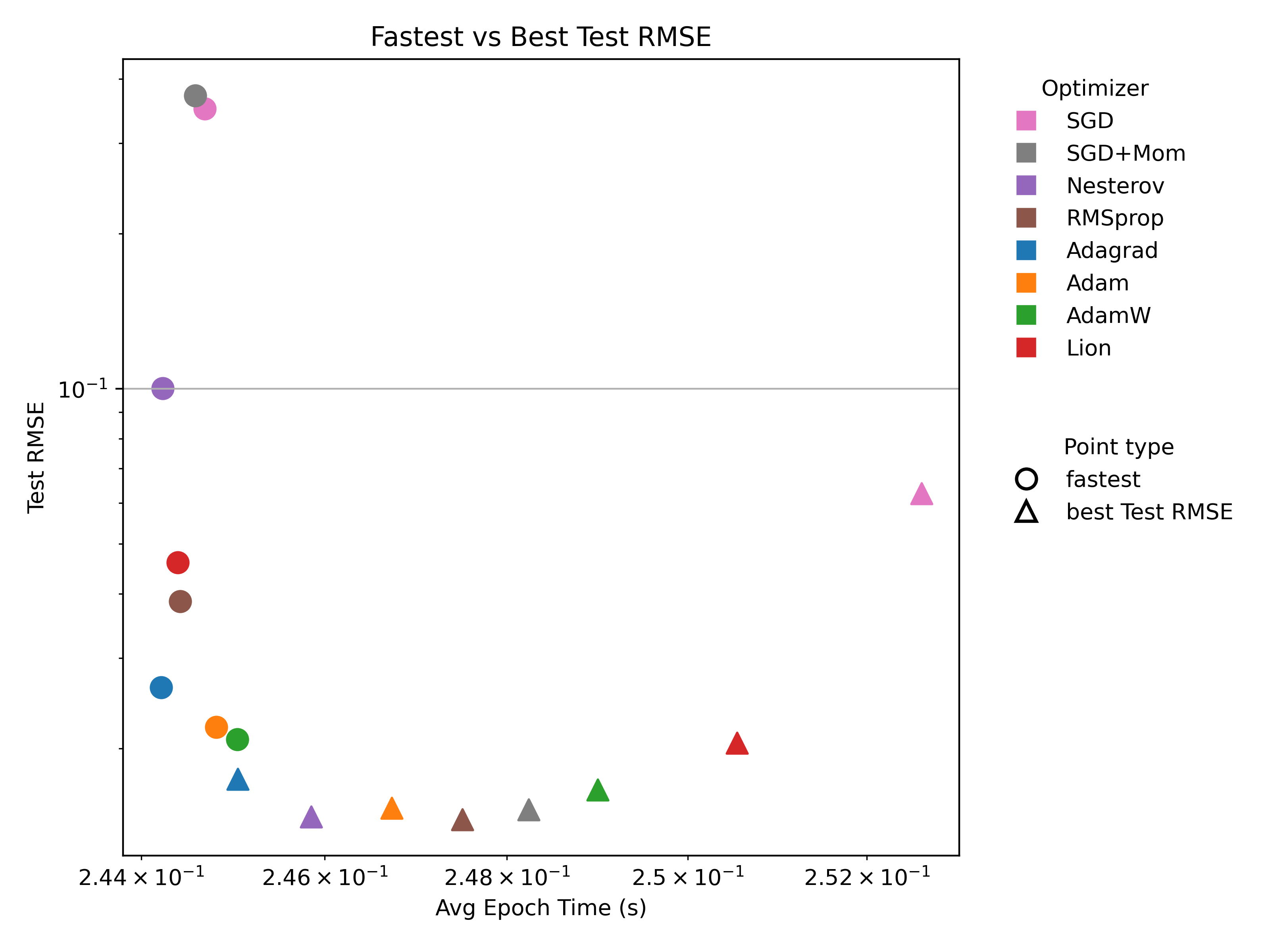}
    \caption{Baseline Optimizers: Fastest vs. Best Test RMSE}
    \label{fig:base-opt-best-RMSE}
\end{figure}

\begin{figure}[h!]
    \centering
    \includegraphics[scale=0.5]{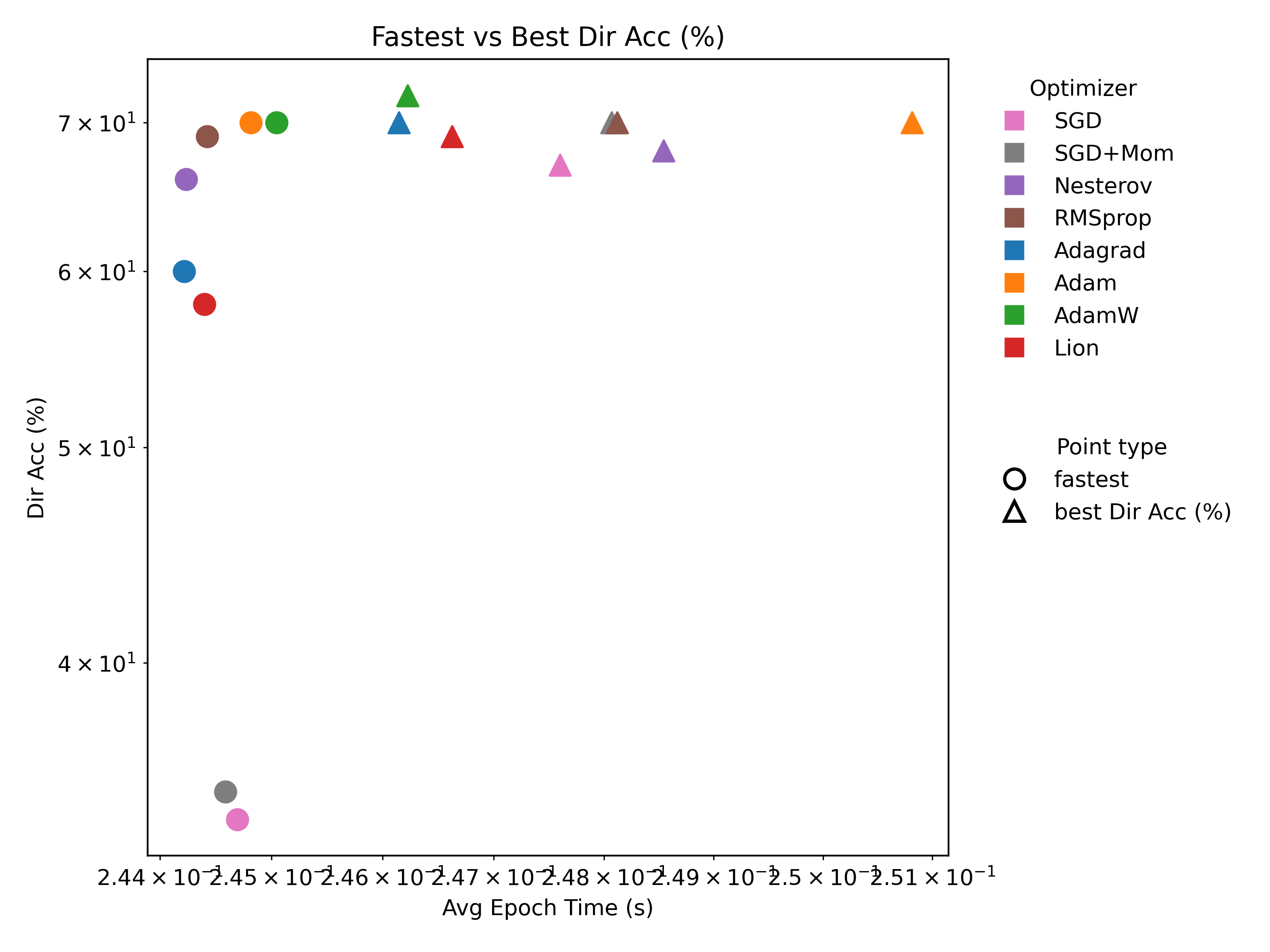}
    \caption{Baseline Optimizers: Fastest vs. Best Directional Accuracy}
    \label{fig:base-opt-best-dir-acc}
\end{figure}

\raggedbottom

\end{document}